\documentclass[12pt,aps,prl]{revtex4}
\usepackage{epsfig}
\usepackage{amsmath}

\normalsize

\def\bA{{\bf A}}

\def\b0{{\bf 0}}

\def\Im{{\rm Im}}

\def\alf{\alpha}
\def\eps{\epsilon}

\def\om{\omega}

\begin{document}

%%%%%%%%%%%%%%%%%%%%%%%%%%%%%%%%% TITLE PAGE %%%%%%%%%%%%%%%%%%%%%%%%%%%%

\title{\large Semiclassical theory of electron drag
 in strong magnetic fields}

\author{Sergej Brener and Walter Metzner}
\affiliation{Max-Planck-Institut f\"ur Festk\"orperforschung, 
  D-70569 Stuttgart, Germany}

\date{\today}

%\maketitle

%%%%%%%%%%%%%%%%%%%%%%%%%%%%%%%%% ABSTRACT %%%%%%%%%%%%%%%%%%%%%%%%%%%%%%

\begin{abstract}
We present a semiclassical theory for electron drag between two 
parallel two-dimensional electron systems in a strong magnetic field, 
which provides a trans\-parent picture of the most salient qualitative 
features of anomalous drag phenomena observed in recent experiments,
especially the striking sign reversal of drag at mis\-matched densities.
The sign of the drag is determined by the curvature of the effective
dispersion relation obeyed by the drift motion of the electrons in a 
smooth disorder potential.
Localization plays a role in explaining activated low temperature
behavior, but is not crucial for anomalous drag {\em per se}. \\
\noindent
\mbox{PACS: 73.21.-b, 73.63.-b, 73.43.-f} \\
\end{abstract}

\maketitle

%%%%%%%%%%%%%%%%%%%%%%%%%%%%%%%%% PAPER %%%%%%%%%%%%%%%%%%%%%%%%%%%%%%%%

%%% Introduction

Electron drag in double-layer two-dimensional electron systems has 
been established as a valuable probe of the electronic state within 
each layer, and also of interlayer interactions \cite{gramila91}.
In a drag experiment a current is driven through one of the layers,
which, via interlayer scattering, produces a drag voltage in the 
other layer. Usually the drag is positive in the sense that 
the drag and drive currents have the same direction, which 
leads to a compensating voltage opposite to the drive current.
In the absence of a magnetic field, drag phenomena in 2D electron 
systems are well understood \cite{rojo99}.
In that case the drag signal can be expressed in terms of the
density response functions of each single layer, as long as the 
interlayer coupling is weak \cite{jauho93}.

Drag experiments have also been performed in the presence of 
strong magnetic fields, where the formation of Landau levels plays 
a crucial role.
Pronounced minima in the drag signal were observed at odd integer 
filling factors for magnetic fields far below the regime where 
spin splitting of Landau levels is seen in the intralayer 
resistivity \cite{hill96,rubel97}.
Completely unexpected was the discovery of a {\em negative}\/ drag 
signal in case of a suitably chosen density mismatch between the 
two layers \cite{feng98,lok01}.
Previous theoretical descriptions \cite{bonsager96,rojo99} 
would predict positive drag at any filling factor.
In more recent work a mechanism for a sign change of drag in 
strong magnetic fields was found \cite{oppen01}, yielding however 
negative drag for equal densities in the two layers, unlike the 
experimental situation.

New hints and constraints for a theory result from a recent
detailed experimental analysis of the temperature dependence of
the drag resistivity by Muraki et al.\ \cite{muraki04}.
At high temperatures, $k_B T \gg \hbar \om_c$, where $\om_c$ 
is the cyclotron frequency, the drag resistivity $\rho_D$ 
follows the conventional $T^2$ behavior.
It decreases with decreasing $T$ down to scales well below 
$\hbar \om_c$, but then, in the data shown for 
$\nu = 2n + \frac{3}{2}$ with $n = 2,3,4,5$, rises again to form 
a pronounced peak at temperatures one order of magnitude below 
$\hbar \om_c$.
At still lower $T$ the drag resistivity decreases rapidly,
consistent with exponentially activated behavior, 
$\rho_D \propto e^{-\Delta/(k_B T)}$.
The $\nu$-dependence of the activation energy $\Delta$ oscillates 
with minima near half-integer filling factors. 
The longitudinal intralayer resistivity $\rho_{xx}$ also exhibits
activated behavior at the lowest temperature, with an activation 
energy matching with $\Delta$ for those $\nu$ where both gaps
could be extracted from the data.
For a density mismatch $\delta\nu = 1$ between the two layers
a low temperature peak with the same shape as for equal densities, 
but with opposite (that is negative) sign was observed. 

Muraki et al.\ \cite{muraki04} interpreted their data in terms of
particle-hole asymmetry and disorder induced localization.
Landau levels are broadened by disorder, leading thus to a 
band of energies for each level. 
Anomalous drag is observed for parameters where the Landau level 
broadening is much smaller than $\hbar\om_c$.
The longitudinal resistivity $\rho_{xx}$ of each layer as well
as the drag resistivity $\rho_D$ is due to electrons in extended 
states of the highest occupied Landau level. 
If the states with energies near the Fermi level $\eps_F$ are 
localized, charge carriers in extended states can be created 
only by thermal activation or scattering across the energy gap 
between $\eps_F$ and the nearest extended state energy level. 
This explains the observation of activated behavior of $\rho_{xx}$ 
and $\rho_D$ at low $T$.
Muraki et al.\ \cite{muraki04} also realized that odd integer
filling factors lead to a particle-hole symmetric occupation
of the partially filled Landau level.
Since particle-hole asymmetry is known to be crucial for drag
in the absence of a magnetic field \cite{kamenev95}, 
it is indeed natural to relate the minima in $\rho_D$ observed 
at odd integer $\nu$ to this symmetry.
Judging from the experiments a necessary requirement for negative
drag seems to be that the partially occupied Landau level in one 
layer be more, the other less than half-filled.
This was noticed already by Feng et al.\ \cite{feng98} and
led them to speculate that electrons in a more than half-filled 
Landau level should acquire a hole-like dispersion relation due 
to disorder. 

A comprehensive transport theory for the intralayer and drag 
resistivity in a strong magnetic field, which captures 
localization and extended states, is not available. 
In this work we present a simple semiclassical picture which 
explains why Landau quantized electrons moving in a smooth disorder
potential behave effectively like band electrons with an electron-like 
dispersion in the lower and a hole-like dispersion in the upper half 
of a disorder broadened Landau level, 
and how this leads to the observed anomalous drag.

Let us first consider the relevant length and energy scales in 
the double-layer system. The distance between the layers varies
from 30 to 120 {\rm nm}. 
The disorder is due to remote donors, which leads to a 
smooth disorder potential with a correlation length $\xi$ of 
order $50-100 \, {\rm nm}$.
The Landau level broadening, which is related to the amplitude
of the disorder potential, has been estimated from low-field 
Shubnikov-de-Haas oscillations to lie in the range 
0.5-2 K \cite{feng98,lok02}.
The magnetic fields at which drag minima at odd integer filling 
and/or negative drag at mismatched densities are observed vary 
over a relatively wide range from $0.1 \, {\rm T}$ to 
$1 \, {\rm T}$ for the cleanest samples, and up to $5 \, {\rm T}$ 
for samples with a slightly reduced mobility.
This corresponds to magnetic lengths $l_B$ between $15$ and 
$80 \, {\rm nm}$.
Hence, $\xi$ and $l_B$ are generally of the same order 
of magnitude, which makes a quantitative theoretical analysis 
rather difficult. 
Since the anomalous drag phenomena are observed over a wide
range of fields, one may hope that qualitative insight can be
gained also by analysing limiting cases. 
For $\xi \ll l_B$ a treatment of disorder within the 
self-consistent Born approximation is possible \cite{raikh93}. 
This route was taken most 
recently by Gornyi et al.\ \cite{gornyi04}, who obtained negative 
contributions to the drag resistivity at mismatched densities 
in agreement with experiment. Localization is not captured by
the Born approximation, and consequently the resistivities obey
power-laws rather than activated behavior for $T \to 0$.
For stronger magnetic fields, on the other hand, localization
will become important, and a semiclassical approximation is a 
better starting point, which is the route we take here. 
Applying a criterion derived by Fogler et al.\ \cite{fogler97}
we estimate that (classical) localization may set in already 
at $0.1 \, {\rm T}$.

%%% Qualitative picture

We now discuss the semiclassical picture of electron states 
and drag on a qualitative level, ignoring spin splitting for
simplicity. 
If the disorder potential varies smoothly, electron states 
lie essentially on contours of equal energy.
These contours form closed loops, corresponding to localized
states, except at a single energy $\eps_0$ in the center of 
each Landau level, for which there is a percolating contour 
through the whole system \cite{tsukuda76}.
If the Landau level broadening induced by the disorder potential 
and also $k_B T$ is much smaller than $\hbar\om_c$, as is the
case in the anomalous drag regime, all Landau levels except one 
are either fully occupied in the bulk of the whole sample or 
completely empty.
At zero temperature and for $\eps_F < \eps_0$ the sample then 
consists of islands where the highest (partially) occupied Landau 
level is locally full, while it is locally empty in the rest of
the sample; for $\eps_F > \eps_0$ the empty regions form 
islands.
In reality, the percolating contour at the center of the Landau
level is broadened to a percolation region for two reasons.
First, electrons near the percolating contour can hop across 
saddlepoints from one closed loop to another, by using for 
a moment some of their cyclotron energy \cite{fogler97}.
Second, electrons near the center of the Landau level can screen 
the disorder potential, such that the percolating equipotential 
line at $\eps_0$ broadens to form equipotential 
terraces \cite{cooper93}.
Hence, there is a region around the percolating path, where 
states are extended.

The percolating region forms a two-dimensional random network 
consisting of links and crossing points \cite{chalker88}.
For simplicity we assume that the links are straight lines. 
The region near the link (except near the end points)
can be parametrized by a cartesian 
coordinate system with a variable $x$ following the equipotential 
lines parallel to the link and $y$ for the transverse direction.
The disorder potential varies essentially only in transverse 
direction, and can thus be represented by a function $U(y)$. 
In the Landau gauge $\bA = (-By,0)$ the Hamiltonian for 
electrons on the link is then translation invariant in 
$x$-direction and the Landau states ($\hbar = 1$ from now on)
\begin{equation}
 \psi_{nk}(x,y) = C_n \, e^{-(y - l_B^2 k)^2/(2l_B^2)} \,
 H_n[(y - l_B^2 k)/l_B] \, e^{ikx}
\end{equation}
are accurate solutions of the stationary Schr\"odinger equation.
Here $H_n$ is the n-th Hermite polynomial and $C_n$ a 
normalization constant.
The corresponding energy is simply
\begin{equation}
 \eps_{nk} = \left( n + {\textstyle\frac{1}{2}} \right) \om_c +
 U(l_B^2 k) \; .
\end{equation}
In the following we drop the Landau level index $n$, because
only the highest occupied level is relevant.
The quantum number $k$ is the momentum associated
with the translation invariance in $x$-direction. It is
proportional to a transverse shift $y_0 = l_B^2 k$ of the
wave function.
The potential $U$ lifts the degeneracy of the Landau levels
and makes energies depend on the momentum along the link. 
The group velocity
\begin{equation}
 v_k = \frac{d\eps_k}{dk} = l_B^2 \, U'(y_0)
\end{equation}
corresponds to the classical drift velocity of an electron in 
crossed electric and magnetic fields.
In our simplified straight link approximation $U'(y_0)$ does not 
depend on $x$. In general it will depend slowly on the longitudinal
coordinate, but near percolating paths away from crossing points 
it has a fixed sign, and hence the group velocity a fixed 
direction, that is the motion on the links is chiral.

The drag between two parallel layers is dominated by electrons 
in the percolating region, corresponding to states near the 
center of the Landau level, since electrons in deeper localized 
states are not easily dragged along. 
In the network picture, macroscopic drag arises as a sum of 
contributions from interlayer scattering processes between
electrons in different links.
If no current is imposed in the drive layer, both layers are
in thermal equilibrium and the currents on the various links
cancel each other on average. 
Now assume that a small finite current is switched on in the
drive layer such that the electrons move predominantly in the
direction of the positive x-axis (the electric current moving
in the opposite direction). This means that the current on
links oriented in the positive x-direction is typically larger
than the current on links oriented in the negative x-direction.
Interlayer scattering processes lead to momentum transfers
between electrons in the drive and drag layer. The preferred
direction of momentum transfers is such that the scattering
processes tend to reduce the current in the drive layer, that
is the interlayer interaction leads to friction. 
Now the crucial point is that electrons moving in the disorder 
potential are not necessarily accelerated by gaining momentum
in the direction of their motion, or slowed down by loosing
momentum (as for free electrons). The fastest electrons are
those near the center of the Landau level: they have the 
highest group velocity on the links of the percolation network
and they get most easily across the saddlepoints.
Electrons in states below the Landau level center are thus
accelerated by gaining some extra momentum in the direction of
their motion, but electrons with an energy above $\eps_0$ are
pushed to still higher energy by adding momentum and are thus
slowed down. 
To understand how negative drag arises, consider the situation
with the highest occupied Landau level of the drive layer less
than, and the one in the drag layer more than half filled.
In that case the electrons in the drag layer receive momentum 
transfers with a predominantly positive x-component.
As a consequence, electrons near the Fermi level of the drag
layer are mostly slowed down if they move on links toward 
positive x-direction, and accelerated if they move on links 
in the negative x-direction, which is the opposite of what
free electrons would do. If no net current is allowed to flow
in the drag layer, an electric field is built up in negative
x-direction which compensates the effective force generated 
by the scattering processes with the drive layer electrons.
The drag signal is thus negative.

%%% Boltzmann equation

To substantiate the above qualitative picture, let us analyze 
the drag between two links using semiclassical transport theory.
We first discuss parallel links in some detail, and then briefly 
the more general case of a finite angle between the links.
For a fixed Landau level index, the electronic states in each
link are fully labelled by their momentum $k$. 
The (non-equilibrium) occupation of states in the links is 
described by a distribution function $f_{\alf}(k)$, where 
$\alf=1,2$ labels the two layers.
We consider the experimental standard setup where a small current 
flows through the ''drive-layer'' ($\alf=2$), generating a 
compensating drag voltage in the ''drag-layer'' ($\alf=1$). 
No net current is allowed to flow in the drag-layer.
Under these conditions, the linear response of the drag layer
to the drive current is determined by the linearized Boltzmann 
equation
\begin{equation}
 \dot k_1 \cdot \frac{\partial f^0_1}{\partial k_1} =
 \left[ \frac{\partial f_1}{\partial t} \right]_{\rm coll}
\end{equation}
with $\dot k_1 = -e E_1$, where $E_1$ is the electric field
leading to the drag voltage.
The interlayer collision term is given by
\begin{eqnarray}
 \left[ \frac{\partial f_1}{\partial t} \right]_{\rm coll}^{12} 
 &=&
 - \int \frac{d k_2}{2\pi} \int \frac{d k'_1}{2\pi} \,
 W^{12}_{k_1,k_2;k'_1,k'_2} \, 
 [\psi_1(k_1) + \psi_2(k_2) - 
  \psi_1(k'_1) - \psi_2(k'_2)]
 \nonumber \\[2mm] && 
 f^0_1(k_1) f^0_2(k_2) 
 [1 - f^0_1(k'_1)] [1 - f^0_2(k'_2)]
 \, 
 \delta(\eps_{k_1} + \eps_{k_2} - \eps_{k'_1} - \eps_{k'_2})
 \; ,
\end{eqnarray}
where $W^{12}_{k_1,k_2;k'_1,k'_2}$ is the rate for a single
interlayer scattering event $k_1 \to k'_1$, $k_2 \to k'_2$ and
the deviation (from equilibrium) functions $\psi_{\alf}$ are 
defined as usually by $f - f^0 = f^0 (1-f^0) \, \psi$.
There is also an intralayer scattering term with a similar
structure.
The distribution function in the drive layer is obtained
from the single layer Boltzmann equation as
$\, \psi_2(k) = - \frac{\pi}{e v_{F2} T} \, j_2 \, v_k \,$,
where $j_2$ is the drive current and $v_{F2}$ the equilibrium
Fermi velocity.
For weak interlayer coupling, the deviation function $\psi_1$
can be neglected in the interlayer collision term, because
it would yield a contribution of order $(W^{12})^2$. 
To determine the relation between the drive current $j_2$ 
and the drag field $E_1$, we multiply the Boltzmann equation
by $v_{k_1}$ and integrate over $k_1$.
The left hand side yields
$-e \int \frac{dk_1}{2\pi} \, {f_1^0}'(\eps_{k_1}) \,
 v_{k_1}^2 E_1 \,$ which tends to 
$\frac{e}{2\pi} \, v_{F1} \, E_1$ at low temperatures, where
$v_{F1}$ is the velocity at the Fermi level of the drive layer.
Using the antisymmetry of the integrand of Eq.\ (5) under
exchange of $k_1$ and $k'_1$, 
the right hand side can be written as
\begin{eqnarray}
 && - \frac{\pi j_2}{e v_{F2} T} \int \frac{d k_1}{2\pi}
 \int \frac{d k_2}{2\pi} \int \frac{d q}{2\pi} \,
 W^{12}_{k_1,k_2;k_1+q,k_2-q} \, 
 (v_{k_2} - v_{k_2-q}) \, 
 (v_{k_1} - v_{k_1+q})
 \nonumber \\[2mm] && \; \times 
 \frac{[f^0_1(\eps_{k_1}) - f^0_1(\eps_{k_1+q})] \,
  [f^0_2(\eps_{k_2}) - f^0_2(\eps_{k_2-q})]}
 {4\sinh^2[(\eps_{k_1+q}-\eps_{k_1})/2T]} \,
 \delta(\eps_{k_1} + \eps_{k_2} - 
 \eps_{k_1+q} - \eps_{k_2-q}) \; .
\end{eqnarray}
The integrated intralayer scattering contributions cancel due 
to the condition of vanishing drag current.
Note that the above integral shares several features
with the general expression for the drag response function, 
as obtained from the Kubo formula \cite{kamenev95,flensberg95}. 
In particular it is symmetric in the layer indices, a property
that depends crucially on the correct form of $\psi_2(k)$.
For a simplified discussion of the most important points we assume 
that the interlayer scattering rate $W^{12}$ depends only on 
momentum transfers $q$ and energy transfers 
$\om = \eps_{k_1+q} - \eps_{k_1}$, and that momentum transfers 
are so small that one can approximate 
$v_{k+q} - v_{k}$ by $q \, \frac{d v_k}{dk}$.
The drag resistivity $\rho_D = - E_1/j_2$ can then be written as
\begin{equation}
 \rho_D = \frac{1}{2\pi e^2} \, 
 \frac{1}{v_{F1} v_{F2}} \, \frac{1}{m_1 m_2}
 \int_0^{\infty} \! dq \, q^2 
 \int_{-\infty}^{\infty}
 \frac{d\om/T}{\sinh^2(\om/2T)} \, W^{12}(q,\om) \,
 \Im\chi_1(q,\om) \, \Im\chi_2(q,\om) \; ,
 \label{rhod}
\end{equation}
where $\chi_{\alf}(q,\om)$ is the dynamical density correlation 
function in layer $\alf$, and the effective masses are given by 
the \emph{curvature} of the dispersion relations at the Fermi
level
\begin{equation}
 \frac{1}{m_{\alf}} = 
 \left. \frac{dv_{k\alf}}{dk} \right|_{k_{F\alf}} \; .
\end{equation}
The integral in Eq.\ (7) is always positive. The sign of 
$\rho_D$ is thus given by the sign of the effective masses,
that is by the curvature of the dispersion at the Fermi 
level. Negative drag is obtained when the dispersion in one
layer is electron-like, and hole-like in the other.
The drag vanishes if the Fermi level in one of the layers
is at an inflection point of the dispersion.
For a quadratic dispersion $\eps_k = \frac{k^2}{2m}$ one
has $v_F = k_F/m$ and Eq.\ (7) reduces to a one-dimensional
version of the well-known semiclassical result for drag
between free electrons in two dimensions \cite{rojo99}.
Returning to Eq.\ (6), it is not hard to generalize the above
results on the sign of $\rho_D$ allowing for larger momentum
transfers $q$ and general momentum dependences of $W^{12}$.
Note that in our case of chiral electrons no backscattering is 
possible, unlike the situation in quantum wires \cite{pustilnik03}.

For parallel links, energy and momentum conservation restrict
the allowed scattering processes very strongly.
At low temperatures, this leads to an exponential suppression
of the drag between parallel links. This has nothing
to do with the exponential suppression of drag observed in the 
experiments, since the links are generically not parallel.
For non-parallel links the sum of momenta on the two links is 
no longer conserved in the scattering process.
Hence scattering processes are suppressed much less at low
temperatures. Computing the drag between non-parallel links
from the linearized Boltzmann equation (a straightforward
generalization of the above steps for parallel links) yields 
a quadratic temperature dependence at low $T$.
The momentum transfers in the drag and drive links are however 
still correlated for non-parallel links, especially when the 
angle between the links is not very large,
and the relative sign of the curvature of the dispersion
in drive and drag layer, respectively, determines the sign of
the drag.
The average curvature vanishes for states in the center of the
Landau level, while it is positive for energies below and 
negative for energies above $\eps_0$.
We thus understand the observation of negative drag when
the Landau level in one layer is less than half-filled, and 
more than half-filled in the other.

Spin can be easily included in the above picture. Since the
interlayer interaction is spin independent, one simply has to
sum over the two spin species (up and down) in both drive and
drag layer, taking the (exchange enhanced) Zeeman spin splitting
of the Landau levels into account. If the Fermi level of one
layer lies between the centers of the highest occupied Landau 
levels for up and down spins, respectively, positive and 
negative contributions to the drag partially cancel each other.
The cancellation is complete due to particle-hole symmetry in
the case of odd integer filling, as observed in experiment.

Within our semiclassical picture anomalous drag, especially
negative drag, is suppressed at temperatures above the Landau 
level width, because then electron- and hole-like states 
within the highest occupied level are almost equally populated. 
This agrees with the results from the Born 
approximation \cite{gornyi04}, and also with experiments.
For the low temperature asymptotics of the drag, the semiclassical
theory yields two different types of behavior, depending on the
filling. 
If the Fermi level does not hit any extended states (for either 
spin species), the drag should vanish exponentially for $T \to 0$,
since thermal activation or scattering of electrons into 
extended states is then suppressed by an energy gap.
By contrast, for a Fermi level within the extended states band
(for at least one spin species) the gap vanishes and the drag 
obeys generally quadratic low temperature behavior, as obtained 
for the drag between non-parallel links. 
Within the Born approximation no localization occurs and the drag 
resistance always vanishes quadratically in the low temperature
limit \cite{gornyi04}.
In high mobility samples localization is negligible at low magnetic 
fields, while an increasing amount of states gets localized at 
higher fields \cite{fogler97}.

In summary, we have presented a semiclassical theory for electron
drag between two parallel two-dimensional electron systems in a 
strong magnetic field, which provides a transparent picture of 
the most salient qualitative features of anomalous drag phenomena 
observed in recent experiments \cite{feng98,lok01,muraki04}.
Localization plays a role in explaining activated low temperature
behavior, but is not crucial for anomalous (especially negative)
drag per se.
A quantitative theory of drag which covers the whole range from
low magnetic fields, where the Born approximation is 
valid \cite{raikh93,gornyi04}. to high fields, where localization
becomes important, remains an important challenge for work to
be done in the future.

\vskip 1cm

{\bf Acknowledgements:} We gratefully acknowledge several important 
discussions with Leonid Glazman, in particular on transport and
drag in one-dimensional channels. He led us to the Boltzmann
equation analysis of drag between links, which substantiated
our semiclassical picture considerably. Special thanks go also
to Rolf Gerhardts for his help in the early stages of this work.
We are also grateful for valuable discussions with E.\ Brener,
W.\ Dietsche, I.\ Gornyi, K.\ von Klitzing, K.\ Muraki, and 
especially Sjoerd Lok.

\vskip 1cm

%%%%%%%%%%%%%%%%%%%%%%%%% REFERENCES %%%%%%%%%%%%%%%%%%%%%%%%%%%%%%%%%%%%%

\vfill\eject

\end{document}